\documentclass[12pt]{iopart}

\usepackage{epsf}
\usepackage{graphicx}
\newcommand{\boldvec}[1]{{\mbox{\boldmath$#1$}}}

\begin{document}

\title[Feshbach resonances and collapsing Bose-Einstein
condensates]{Feshbach resonances and collapsing Bose-Einstein
  condensates}

\author{J. N. Milstein$^1$, C. Menotti$^{2,3}$, and M. J. Holland$^1$}

\address{$^1$JILA and Department of Physics, University of
Colorado at Boulder, CO 80309-0440}

\address{$^2$Dipartimento di Fisica, Universit\`a di Trento and
  BEC-INFM, I-38050 Povo, Italy}

\address{$^3$Dipartimento di Matematica e Fisica,  
Universit\`{a} Cattolica del Sacro Cuore,  
I-25121 Brescia, Italy}

\ead{murray.holland@colorado.edu}

\begin{abstract}
  We investigate the quantum state of burst atoms seen in the recent
  Rb-85 experiments at JILA. We show that the presence of a resonance
  scattering state can lead to a pairing instability generating an
  outflow of atoms with energy comparable to that observed. A
  resonance effective field theory is used to study this dynamical
  process in an inhomogeneous system with spherical symmetry.
\end{abstract}
\maketitle


\section{Introduction}

The ability to dynamically modify the nature of the microscopic
interactions in a Bose-Einstein condensate---an ability virtually
unique to the field of dilute gases---opens the way to the exploration
of a range of fundamental phenomena. A striking example of this is the
Bosenova experiment carried out in the Wieman group at
JILA~\cite{donley} which explored the mechanical collapse instability
arising from an attractive interaction.  This collapse resulted in an
unanticipated burst of atoms, the nature of which is a subject of
current debate. In this article we suggest a possible mechanism for
the formation of these bursts by application of an effective quantum
field theory which includes explicitly the resonance scattering
physics.

The Bosenova experiment conducted by the JILA group consisted of the
following elements. A conventional stable Bose-Einstein condensate was
created in equilibrium. The group then utilized a Feshbach resonance
to abruptly switch the interactions to be attractive inducing an
implosion.  One might have predicted that the rapid increase in
density would simply lead to a rapid loss of atoms, primarily through
inelastic three-body collisions. In contrast, what was observed was
the formation of an energetic burst of atoms emerging from the
implosion.  Although the energy of these atoms was much larger than
that of the condensate, the energy was insignificant when compared to
the molecular binding energy which characterizes the energy released
in a three-body collision.  In the end, what remained was a remnant
condensate which appeared distorted and was believed to be in a highly
excited collective state.

One theoretical method which has been extensively explored to explain
this behavior has been the inclusion of a decay term into the
Gross-Pitaevskii equation as a way to account for the atom
loss~\cite{duine,santos,saito,savage}.  Aside from its physical
application to the Bosenova problem, the inclusion of three-body loss
as a phenomenological mechanism represents an important mathematical
problem, since the nonlinear Schr\"odinger equation allows for a class
of self-similar solutions in the unstable regime.  The local collapses
predicted in this framework can generate an outflow even within this
zero temperature theory. However, there are a number of aspects which
one should consider when applying the extended Gross-Pitaevskii
equation to account for the observations made in the JILA experiment.

\begin{figure}
\begin{center}\
\epsfysize=80mm
\epsfbox{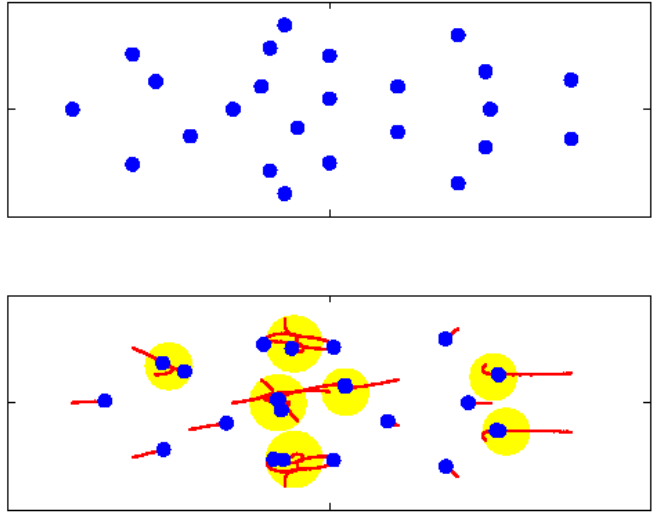}
\end{center}
\caption{\label{correlations}
  A collapsing classical system with point-like objects interacting
  via a Gaussian attractive potential. The top graph shows the initial
  configuration. The lower graph displays the evolved state
  demonstrating clustering of the particles (illustrated as yellow
  regions). The red lines mark the particle trajectories.}
\end{figure}

The first problematic issue is the potential breakdown of the
principle of attenuation of correlations. This principle is essential
in any quantum or classical kinetic theory as it allows multiparticle
correlations to be factorized. This assumption is especially evident
in the derivation of the Gross-Pitaevskii equation where all explicit
multiparticle correlations are dropped. However, as shown in
Figure~\ref{correlations}, even a simple classical model may exhibit
clustering when mechanically unstable which appears to invalidate the
assumption of an attenuation of correlations. Furthermore, there is
also considerable evidence for this instability toward pair formation
in the mechanically unstable quantum theory~\cite{jeon}.

A second difficulty with motivating the Gross-Pitaevskii approach is
that by this method one describes the interactions as energy
independent through a single parameter, the scattering length, which
is determined from the $s$-wave scattering phase shift at zero
scattering energy. Near a Feshbach resonance, the proximity of a bound
state in a closed potential to the zero of the scattering continuum
can lead to a strong energy dependence of the scattering. Exactly on
resonance, the $s$-wave scattering length passes through infinity and,
in this situation, the Gross-Pitaevskii equation is undefined.

These two fundamental difficulties with the Gross-Pitaevskii approach
led us to reconsider this problem. We were motivated by the fact that
the same experimental group at JILA recently performed a complementary
experiment~\cite{donley2} which provided key insights into the
Bosenova system. What was remarkable in these new experiments was
that, even with a large positive scattering length in which the
interactions were repulsive, a burst of atoms and a remnant
condensate were observed. Furthermore, in the large positive
scattering length case, simple effective field theories which included
an explicit description of the Feshbach resonance physics were able to
provide an accurate quantitative comparison with the data
\cite{kokkelmans1,mackie,kohler}. The theory showed the burst to arise
from the complex dynamics of the atom condensate coupled to a coherent
field of exotic molecular dimers of a remarkable physical size and
near the threshold binding energy.

In this paper, we draw the connections between the two JILA
experiments. We pose and resolve the question as to whether the burst
of atoms in the Bosenova collapse could arise in a similar way as in
the Ramsey fringe experiment---from the formation of a coherent
molecular superfluid. This hypothesis is tested by applying an
effective field theory for resonance superfluidity to the collapse.
For fermions, the case of resonance superfluidity in an inhomogeneous
system has been treated in the local density approximation, using
essentially the uniform solution at each point in space
\cite{chiofalo}.  For the collapse of a Bose-Einstein condensate, as
we wish to treat here, the local density approximation is not valid,
and the calculation must be performed on a truly inhomogeneous system.
This represents the first time that the resonance superfluidity theory
has been applied to a system of this type.

\section{Effective Field Theory}

\begin{figure}
\begin{center}\
\epsfysize=60mm
\epsfbox{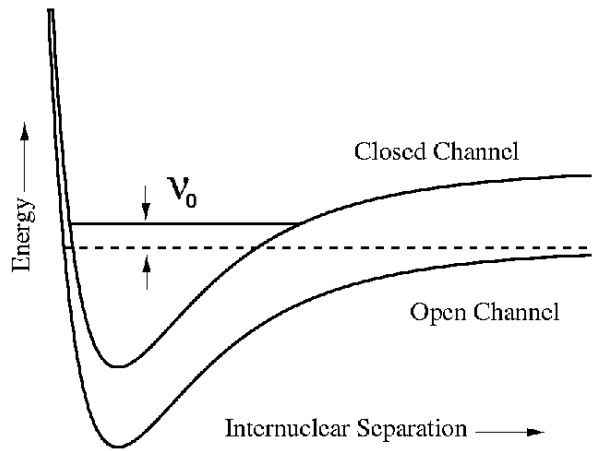}
\end{center}
\caption{\label{feshbach}
  A Feshbach resonance results when a closed channel potential
  possesses a bound state in proximity to the scattering energy in an
  open channel potential. The detuning of the bound state from the
  edge of the collision continuum is denoted by $\nu_0$.  }
\end{figure}

In the Feshbach resonance illustrated in Figure~\ref{feshbach}, the
properties of the collision of two ground state atoms is controlled
through their resonant coupling to a bound state in a closed channel
Born-Oppenheimer potential. By adjusting an external magnetic field,
the scattering length can be tuned to have any value. This field
dependence of the scattering length is characterized by the detuning
$\nu_0$ and obeys a dispersive profile given by $a(\nu_0)=a_{\rm
  bg}(1-\kappa/(2\nu_0))$, with $\kappa$ the resonance width and
$a_{\rm bg}$ the background scattering length. In fact, all the
scattering properties of a Feshbach resonance system are completely
characterized by just three parameters $U_0=4\pi\hbar^2a_{\rm bg}/m$,
$g_0=\sqrt{\kappa U_0}$, and $\nu_0$. Physically, $U_0$ represents the
energy shift per unit density on the single particle eigenvalues due
to the background scattering processes, while $g_0$, which has
dimensions of energy per square-root density, represents the coupling
of the Feshbach resonance between the open and closed channel potentials.

We now proceed to construct a low order many-body theory which
includes this resonance physics. The Hamiltonian for a dilute gas of
scalar bosons with binary interactions is given in complete generality
by
\begin{equation}
\fl
H=\int d^3x\; \psi^{\dag}_a(\boldvec{x}) H_a(\boldvec{x})
\psi_a(\boldvec{x})
+\int d^3xd^3x'\;
\psi^{\dag}_a(\boldvec{x})\psi^{\dag}_a(\boldvec{x'}) V
(\boldvec{x},\boldvec{x'})
\psi_a(\boldvec{x'})\psi_a(\boldvec{x})
\label{hamil}
\end{equation}
where $H_a(\boldvec{x})$ is the single particle Hamiltonian, $V
(\boldvec{x},\boldvec{x'})$ is the binary interaction potential, and
$\psi_a(\boldvec{x})$ is a bosonic scalar field operator. In cold
quantum gases, where the atoms collide at very low energy, we are only
interested in the behavior of the scattering about a small energy
range above zero. There exist many potentials which replicate the low
energy scattering behavior of the true potential; therefore, it is
convenient to carry out the calculation with the simplest one, the
most convenient choice being to take the interaction potential as a
delta-function pseudopotential when possible.

For a Feshbach resonance this choice of pseudopotential is generally
not available since the energy dependence of the scattering implies
that a minimal treatment must at least contain a spread of
wave-numbers which is equivalent to the requirement of a nonlocal
potential. Since the solution of a nonlocal field theory is
inconvenient, we take an alternative but equivalent approach. We
include into the theory an auxiliary molecular field operator
$\psi_m(\boldvec{x})$ which obeys Bose statistics and describes the
collision between atoms in terms of two elementary components: the
background collisions between atoms in the absence of the resonance
interactions and the conversion of atom pairs into molecular states.
This allows us to construct a local field theory with the property
that when the auxiliary field is integrated out, an effective
Hamiltonian of the form given in Eq.~(\ref{hamil}) is recovered with a
potential $V(\boldvec x,\boldvec x')=V(|\boldvec x-\boldvec x'|)$
which generates the form of the two-body $T$-matrix predicted by
Feshbach resonance theory~\cite{kokkelmans2}. The local Hamiltonian
which generates this scattering behavior:
\begin{eqnarray}
H&=&\int d^3x\; \psi^{\dag}_a(\boldvec{x})\Bigl(
-\frac{\hbar^2}{2m}\nabla_x^2+V_a(\boldvec{x})-\mu_a\Bigr)
\psi_a(\boldvec{x})\nonumber\\
&&+\int d^3x\; \psi^{\dag}_m(\boldvec{x})\Bigl(
-\frac{\hbar^2}{4m}\nabla_x^2+V_m(\boldvec{x})-\mu_m\Bigr)
\psi_m(\boldvec{x})\nonumber\\
&&{}+{U\over2}\int d^3x\;
\psi^{\dag}_a(\boldvec{x})\psi^{\dag}_a(\boldvec{x})
\psi_a(\boldvec{x})\psi_a(\boldvec{x})\nonumber\\
&&{}+{g\over2}\int d^3x\;
\psi^{\dag}_m(\boldvec{x})
\psi_a(\boldvec{x})\psi_a(\boldvec{x})
\label{localhamil}
\end{eqnarray}
has the intuitive structure of resonant atom-molecule coupling. Here
$V_{a,m}$ are the external potentials and $\mu_{a,m}$ are the chemical
potentials. The subscripts ${a,m}$ represent the atomic and molecular
contributions, respectively. The Feshbach resonance is controlled by
the magnetic field which is incorporated into the theory by the
detuning $\nu=\mu_m-2\mu_a$ between the atomic and molecular fields.
The Hamiltonian in equation~(\ref{localhamil}) contains the three
parameters $U$, $g$, and $\nu$ which account for the complete
scattering properties of the Feshbach resonance. It is important to
keep in mind that they are distinct from the bare parameters $U_0$,
$g_0$, and $\nu_0$ introduced above. In order for the local
Hamiltonian given in equation~(\ref{localhamil}) to be applicable, one
must introduce into the field theory a renormalized set of parameters
each containing a momentum cutoff associated with a maximum wavenumber
$K$.  This need not be physical in origin but should exceed the
momentum range of occupied quantum states. The relationships between
the renormalized and bare parameters are given by: $U=\Gamma U_0$,
$g=\Gamma g_0$, and $\nu=\nu_0+\alpha gg_0/2$, where
$\alpha=mK/(2\pi^2\hbar^2)$ and $\Gamma=(1-\alpha
U_0)^{-1}$~\cite{kokkelmans2}.  All the results presented here have
been shown to be independent of the momentum cutoff in the theory.

\begin{figure}
\begin{center}\
\epsfysize=50mm
\epsfbox{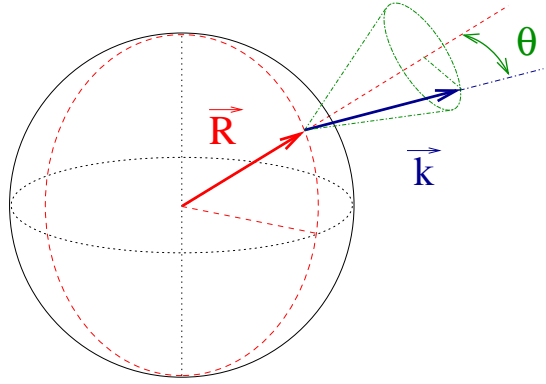}
\end{center}
\caption{\label{Rktheta}
  Illustration of the spherically symmetric geometry used, as defined
  by the center of mass vector $\boldvec R$, relative momentum vector
  $\boldvec k$, and the angle $\theta$ between them.}
\end{figure}

We define the condensates in terms of the mean-fields of the operators
$\phi_a(\boldvec{x})=\bigl<\psi_a(\boldvec{x})\bigr>$ and
$\phi_m(\boldvec{x})=\bigl<\psi_m(\boldvec{x})\bigr>$ along with the
fluctuations about the atomic field
$\chi_a(\boldvec{x})=\psi_a(\boldvec{x})-\phi_a(\boldvec{x})$. Note
that, in principle, there is also a term which involves the
fluctuations about $\phi_m(\boldvec x)$. Assuming the occupation of
$\phi_m(\boldvec x)$ to be small (less than 2\% in the simulations we
present), we drop higher order terms arising from fluctuations about
this mean-field which do not give a significant correction to our
results.  As discussed in reference \cite{kokkelmans2}, we derive four
equations: two corresponding to a Schr\"odinger evolution of the mean
fields
\begin{eqnarray}
  i\hbar{d\phi_a({\boldvec x})\over dt} &=&
  \Bigl(
  -\frac{\hbar^2}{2m}\nabla_x^2+V_a(\boldvec{x})-\mu_a
  +U\bigl[|\phi_a(\boldvec x)|^2+2G_N(\boldvec x,\boldvec x)\bigr]
  \Bigr)\phi_a(\boldvec x) \nonumber\\&&{}+
  \bigl[UG_A(\boldvec x,\boldvec x)+g\phi_m(\boldvec x)\bigr]
  \phi_a^*(\boldvec x)\nonumber\\
  i\hbar{d\phi_m({\boldvec x})\over dt} &=&
  \Bigl(
  -\frac{\hbar^2}{4m}\nabla_x^2+V_m(\boldvec{x})-\mu_m\Bigr)\phi_m(\boldvec x)
  +\frac{g}{2}\bigl[\phi_a^2(\boldvec x)+G_A(\boldvec x,\boldvec x)\bigr]
\label{meanfield}
\end{eqnarray}
and two corresponding to the Louiville space evolution of the normal
density
$G_N(\boldvec{x},\boldvec{x}')=\bigl<\chi_a^{\dag}(\boldvec{x}')
\chi_a(\boldvec{x})\bigr>$ and of the anomalous density
$G_A(\boldvec{x},\boldvec{x}')=\bigl<\chi_a(\boldvec{x}')
\chi_a(\boldvec{x})\bigr>$
\begin{equation}
i\hbar\frac{\partial {\cal G}}{dt}=\Sigma{\cal G}
-{\cal G}\Sigma^{\dag}.
\label{fluctfield}
\end{equation}
The density matrix and self-energy matrix are defined
respectively as
\begin{eqnarray}
{\cal G}(\boldvec x,\boldvec x')&=&
\left(\begin{array}{cc}
\bigl<\chi_a^{\dag}(\boldvec{x}')\chi_a(\boldvec{x})\bigr> &
\bigl<\chi_a(\boldvec{x}')\chi_a(\boldvec{x})\bigr> \\
\bigl<\chi_a^{\dag}(\boldvec{x}')\chi_a^{\dag}(\boldvec{x})\bigr> &
\bigl<\chi_a(\boldvec{x}')\chi_a^{\dag}(\boldvec{x})\bigr>
\end{array}\right)\nonumber\\
\Sigma(\boldvec x,\boldvec x')&=&
\left(\begin{array}{cc}
H(\boldvec x,\boldvec x')&\Delta(\boldvec x,\boldvec x')\\
-\Delta^*(\boldvec x,\boldvec x')&-H^*(\boldvec x,\boldvec x')
\end{array}\right).
\end{eqnarray}
The convenience of choosing a microscopic model in which the potential
couplings are of contact form is now evident since the elements of the
self-energy matrix $\Sigma$ are diagonal in $\boldvec x$ and $\boldvec
x'$ with non-zero elements
\begin{eqnarray}
H(\boldvec x,\boldvec x)&=&
-\frac{\hbar^2}{2m}\nabla_x^2+V_a(\boldvec{x})-\mu_a
  +2U\bigl[|\phi_a(\boldvec x)|^2+G_N(\boldvec x,\boldvec x)\bigr]
  \nonumber\\
  \Delta(\boldvec x,\boldvec x)&=&
  U\bigl[\phi_a^2(\boldvec x)+G_A(\boldvec x,\boldvec x)\bigr]
  +g\phi_m(\boldvec x).
\end{eqnarray}
Since the normal density and anomalous pairing field are both
six-dimensional objects, it is very difficult to solve these equations
in an arbitrary geometry. For this reason we consider the case of
greatest symmetry consisting of a spherical trap. Here we can reduce
the problem to one of only three dimensions, which is still nontrivial
to treat, according to the following procedure. To begin with, it is
convenient to write the elements of the single particle density matrix
in center of mass $\boldvec{R}=(\boldvec{x}+\boldvec{x}')/2$ and
relative $\boldvec{r}=\boldvec{x}-\boldvec{x}'$ coordinates. The
normal density then takes on a familiar structure corresponding to the
Wigner distribution
\begin{eqnarray}
G_N(\boldvec R,\boldvec k)&=&
\int d^3r\,\bigl<\chi_a^{\dag}(\boldvec R-{\boldvec r}/2)
\chi_a(\boldvec R+{\boldvec r}/2)\bigr>e^{
-i{\boldvec{\scriptstyle k}}\cdot{\boldvec{\scriptstyle r}}}
\nonumber\\&=&
\int d^3r\,G_N(\boldvec R,\boldvec r)e^{
-i{\boldvec{\scriptstyle k}}\cdot{\boldvec{\scriptstyle r}}}
\label{fgn}
\end{eqnarray}
which in the high-temperature limit will map on to the particle
distribution function $f(\boldvec{R},\boldvec{k})$ for a classical
gas. Correspondingly, the anomalous density can also be written in
Fourier space as
\begin{eqnarray}
G_A(\boldvec R,\boldvec k)&=&
\int d^3r\,\bigl<\chi_a(\boldvec R-{\boldvec r}/2)
\chi_a(\boldvec R+{\boldvec r}/2)\bigr>e^{
-i{\boldvec{\scriptstyle k}}\cdot{\boldvec{\scriptstyle r}}}
\nonumber\\&=&
\int d^3r\,G_A(\boldvec R,\boldvec r)e^{
-i{\boldvec{\scriptstyle k}}\cdot{\boldvec{\scriptstyle r}}}.
\label{fga}
\end{eqnarray}
In this geometry, the angular dependence of $\boldvec{R}$ is
irrelevant, and the cylindrical symmetry about $\boldvec{R}$ allows
the wavevector $\boldvec{k}$ to be represented by its length and the
one remaining angle as illustrated in Figure~\ref{Rktheta}.
This allows us to represent the density distributions in
three-dimensions as
\begin{equation}
G(\boldvec{R},\boldvec{k})=G(R,k,\theta)
\end{equation}
where $G$ corresponds to either the normal ($G_N$) or anomalous
($G_A$) density. It is now straightforward to rewrite
equations~(\ref{meanfield}) and (\ref{fluctfield}) in this coordinate
system. It is worth pointing out the simple structure of the kinetic
energy contributions to equation~(\ref{fluctfield}) which for the
$G_N$ and $G_A$ components take the corresponding forms respectively:
\begin{eqnarray}
\left(\nabla^2_{x}-\nabla^2_{x'}\right)
G_N(\boldvec x,\boldvec x')&=&
2\left(\nabla_{R}
\cdot\nabla_{r}\right)
G_N(\boldvec R,\boldvec r)
\label{gneqn}
\\
\left(\nabla^2_{x}+\nabla^2_{x'}\right)
G_A(\boldvec x,\boldvec x')&=&
\left(\frac12\nabla^2_{R}
+2\nabla^2_{r}\right)
G_A(\boldvec R,\boldvec r).
\label{gaeqn}
\end{eqnarray}
One may now take the Fourier transform with respect to $\boldvec r$ as
indicated by equations~\ref{fgn} and~\ref{fga}, replacing
$\nabla_r\rightarrow i\boldvec{k}$. The gradient operator $\nabla_R$
can be expressed in any representation, but it is most convenient to
use spherical polar coordinates aligned with the $\boldvec k$
direction vector
\begin{equation} \nabla_R={\hat
    R}\frac{\partial}{\partial R} +{\hat
    \theta}\frac1R\frac{\partial}{\partial\theta}
  +{\hat\varphi}\frac1{\sin\theta}\frac{\partial}{\partial\varphi}
\end{equation}
where $\varphi$ is the azimuthal angle about $\boldvec k$ (which will
eventually drop out in our chosen symmetry), and $\hat R$,
$\hat\theta$, and $\hat\varphi$ are the spherical unit vectors in the
$R$, $\theta$, and $\varphi$ directions, respectively. Noting that
$\hat R\cdot\boldvec k=k\cos\theta$, $\hat\theta\cdot\boldvec k=
-k\sin\theta$, and $\hat\varphi\cdot\boldvec k=0$, we arrive at the
following expression for the differential operator in
equation~\ref{gneqn}
\begin{equation}
\nabla_{R}\cdot\boldvec k
=k\left(\cos\theta\frac{\partial}{\partial R}
-\frac{\sin\theta}{R}\frac{\partial}{\partial\theta}\right).
\end{equation}
Furthermore, the spherical Laplacian for a system with no azimuthal
dependence, as required in equation~\ref{gaeqn}, is given by
\begin{equation}
  \nabla^2_{R}= \frac1{R^2}\frac{\partial}{\partial
    R}\Bigl(R^2\frac{\partial}{\partial R}
  \Bigr)+\frac1{R^2\sin\theta}\frac{\partial}{\partial\theta}
  \Bigl(\sin\theta\frac{\partial}{\partial\theta}\Bigr).
\end{equation}
In practice, we expand the $\theta$ dependence of $G_N$ and $G_A$ in
terms of the orthogonal Legendre polynomials, and the angular
derivatives are then easily implemented via the usual recursion
relations.

\section{Results and Analysis}

\begin{figure}
\begin{center}\
\epsfysize=80mm
\epsfbox{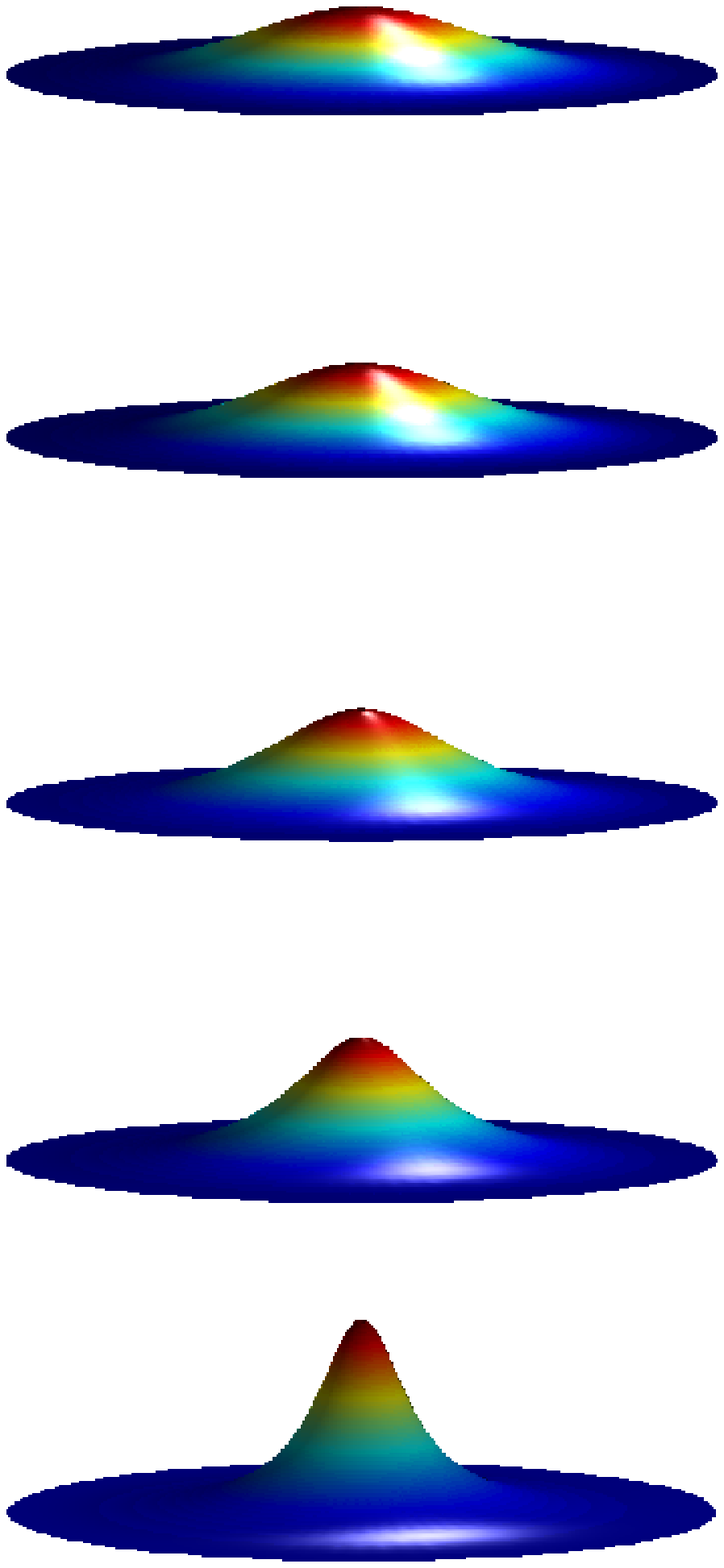}
\hspace*{1cm}
\epsfysize=80mm
\epsfbox{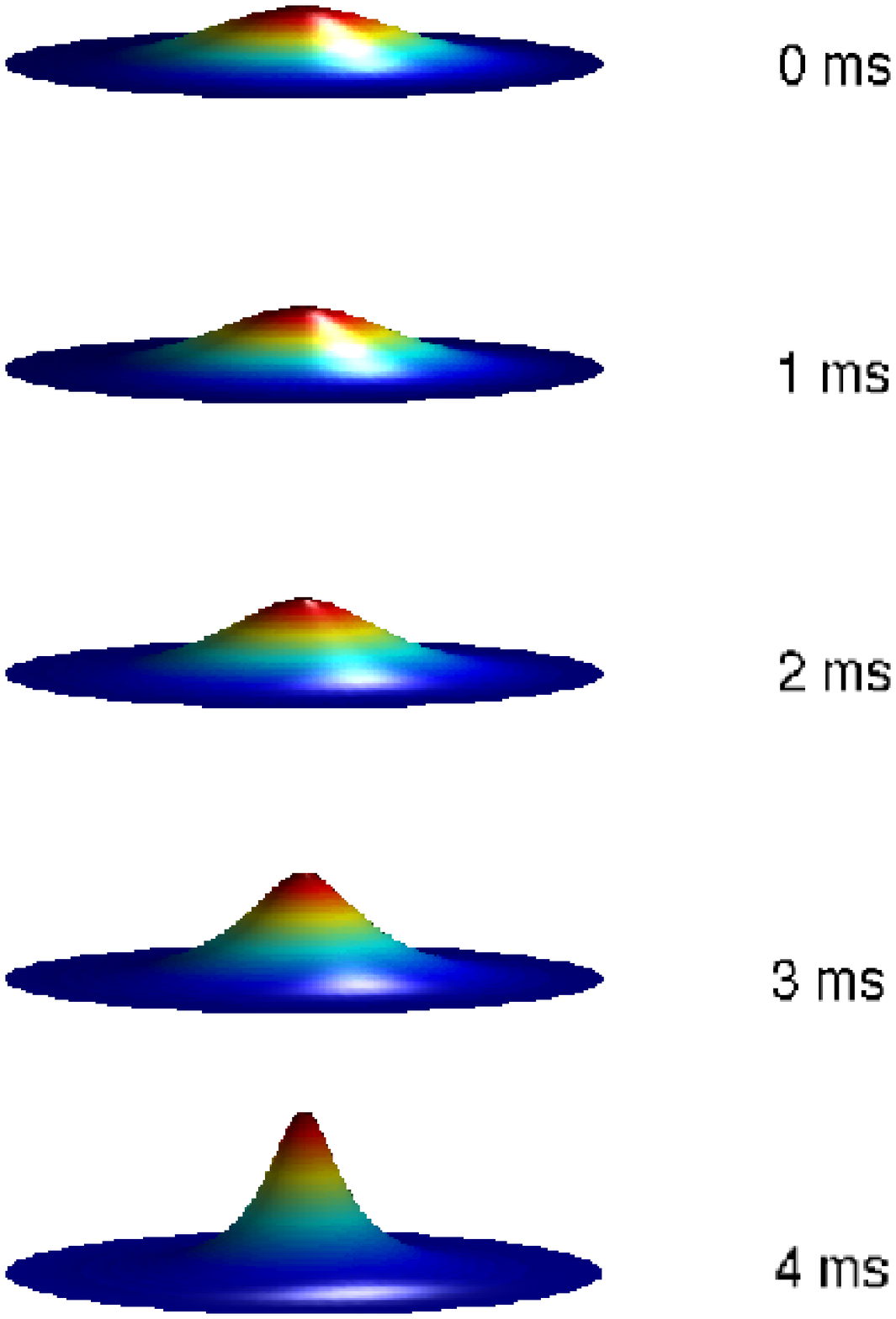}
\end{center}
\caption{\label{gpcollapse}
  A direct comparison of the collapse between the Gross-Pitaevskii
  (left) and the resonance approach (right) within the regime of
  applicability of the Gross-Pitaevskii equation. Each horizontal pair
  is at the same time step with time increasing from top to bottom. As
  expected we observe no appreciable difference between the two
  methods.}
\end{figure}

As an initial test, we expect the resonance theory to give a similar
prediction to the Gross-Pitaevskii equation in the initial phase of
the collapse when the quantum depletion is small.
Figure~\ref{gpcollapse} shows a direct comparison between the
Gross-Pitaevskii approach and the resonance theory. The same initial
conditions were used for all our simulations; 1000 rubidium-85 atoms
in the ground state of a 10 Hz harmonic trap. For all the images we
present, the results of the three-dimensional calculation, in our
spherical geometry, are illustrated as a two-dimensional slice through
the trap center. In the Gross-Pitaevskii solution we used a scattering
length of $-200\,a_0$ where $a_0$ is the Bohr radius.  For comparison,
the Feshbach resonance theory uses a positive background scattering
length of $50\,a_0$ and a resonance width and detuning respectively of
15~kHz and 2.8~kHz. These parameters give the same effective
scattering length as the one used in the Gross-Pitaevskii evolution,
but nowhere in the resonance theory does the effective scattering
length appear explicitly.  As is evident, there is no noticeable
discrepancy between the two approaches over this short timescale.
Eventually we expect these theories to diverge significantly as the
density increases and the coupling between the atomic and molecular
degrees of freedom becomes stronger.  However, at this stage the
agreement is a demonstration that our renormalized theory correctly
allows us to tune the interactions in an inhomogeneous situation.

\begin{figure}
\begin{center}\
\epsfysize=100mm
\epsfbox{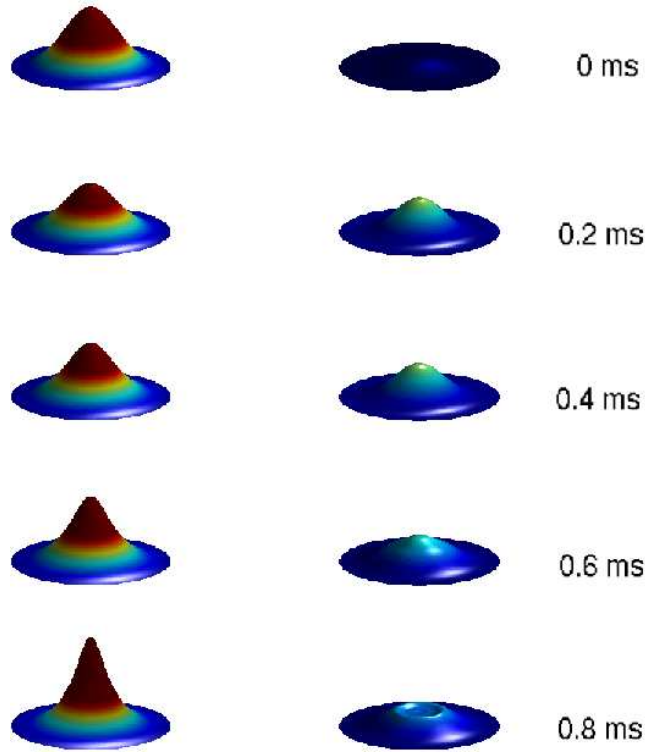}
\end{center}
\caption{\label{MHFBcollapse}
  The simulation of the collapse in the resonance theory showing the
  time evolution of the condensed fraction $\phi_a(\boldvec x)$ (left)
  and noncondensed fraction $G_N(\boldvec x,\boldvec x)$ (right). Each
  horizontal pair is taken at the same instant of time, with time
  increasing from top to bottom. It is evident that non condensate
  atoms are produced during the collapse dynamics, forming rings which
  propagate from the center of the cloud outward.}
\end{figure}

We now proceed to a more complex situation in which the timescales for
the atom-molecule coupling and the collapse dynamics are more
compatible.  From a numerical point of view, it becomes convenient to
increase the resonance width to 1.5~MHz and the detuning to 14~kHz so
that the effect of the atom-molecule coupling will appear in the first
stage of the collapse.  This allows us to form a complete picture of
the dynamics involving the atomic collapse and the simultaneous
coupling to a coherent molecular field. The numerical calculation is
shown as a movie in Fig.~\ref{MHFBcollapse} for both the condensed and
noncondensed components. One sees the formation of a significant
fraction of noncondensed atoms--a feature not described within the
Gross-Pitaevskii framework. During a time evolution of 0.8~ms the
condensate fraction falls to approximately 80\% of its initial value,
while the noncondensate fraction reaches a peak at around 20\%. The
amplitude of the scalar field $\phi_m$ remains below the 2\% level at
all times.

\begin{figure}
\begin{center}\
\epsfysize=70mm
\epsfbox{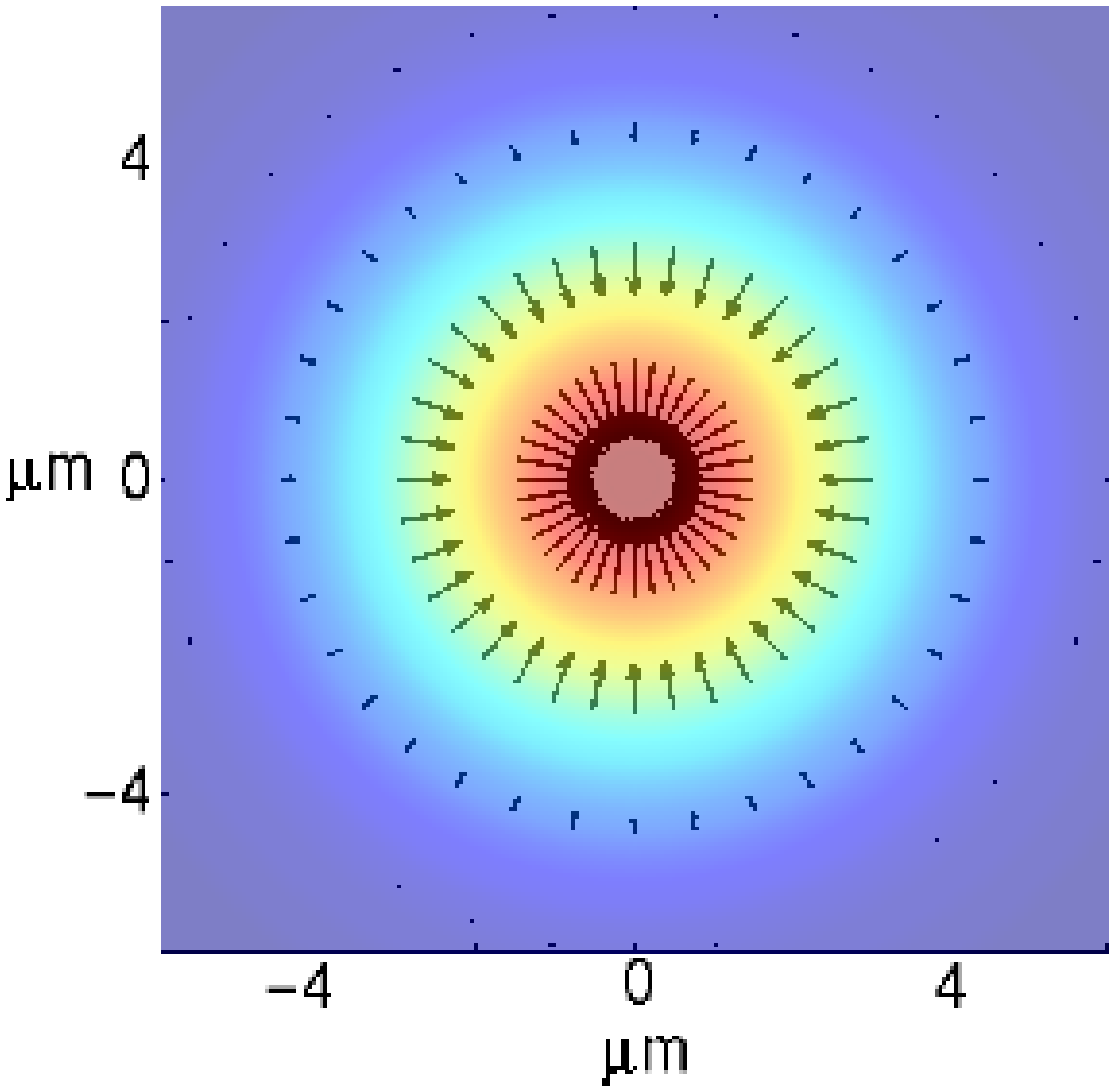}
\epsfysize=70mm
\epsfbox{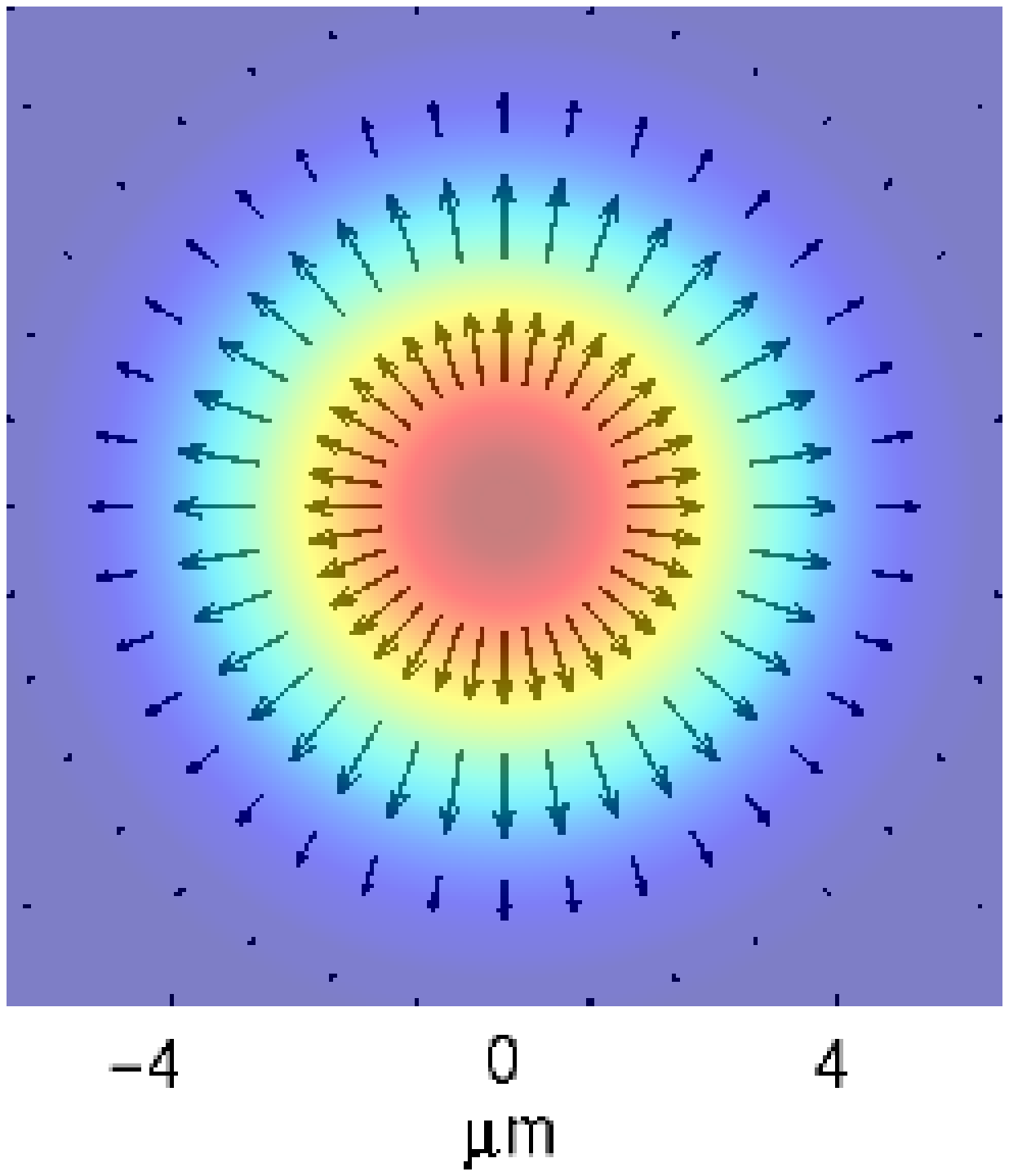}
\end{center}
\caption{\label{gnquiver}
  The velocity fields for the condensate component $\phi_a(\boldvec
  x)$ (left) and the noncondensed component $G_N(\boldvec x,\boldvec
  x)$ (right) midway through the simulation (0.4~ms). The color
  contours indicate the densities and the velocity fields are
  represented in direction and strength by the arrows. This clearly
  shows, that in the resonance theory, as the condensate collapses
  inward, the noncondensate atoms that are generated flow outward.}
\end{figure}

To better illustrate the behavior of the atoms during the collapse we
present the flow of the different distributions involved.  The
condensate velocity field is shown in Figure~\ref{gnquiver}. It
exhibits similar characteristics to those predicted by the
Gross-Pitaevskii theory, which without loss, predicts that the
condensed atoms will always accelerate toward the trap center. In
contrast, the velocity field of the atoms outside of the condensate is
radially outward. The production of this component is quite
interesting because it is this same component which in the theory of
the Ramsey fringe experiment was quantitatively determined to give
rise to the burst.

\begin{figure}
\begin{center}\
\epsfysize=80mm
\epsfbox{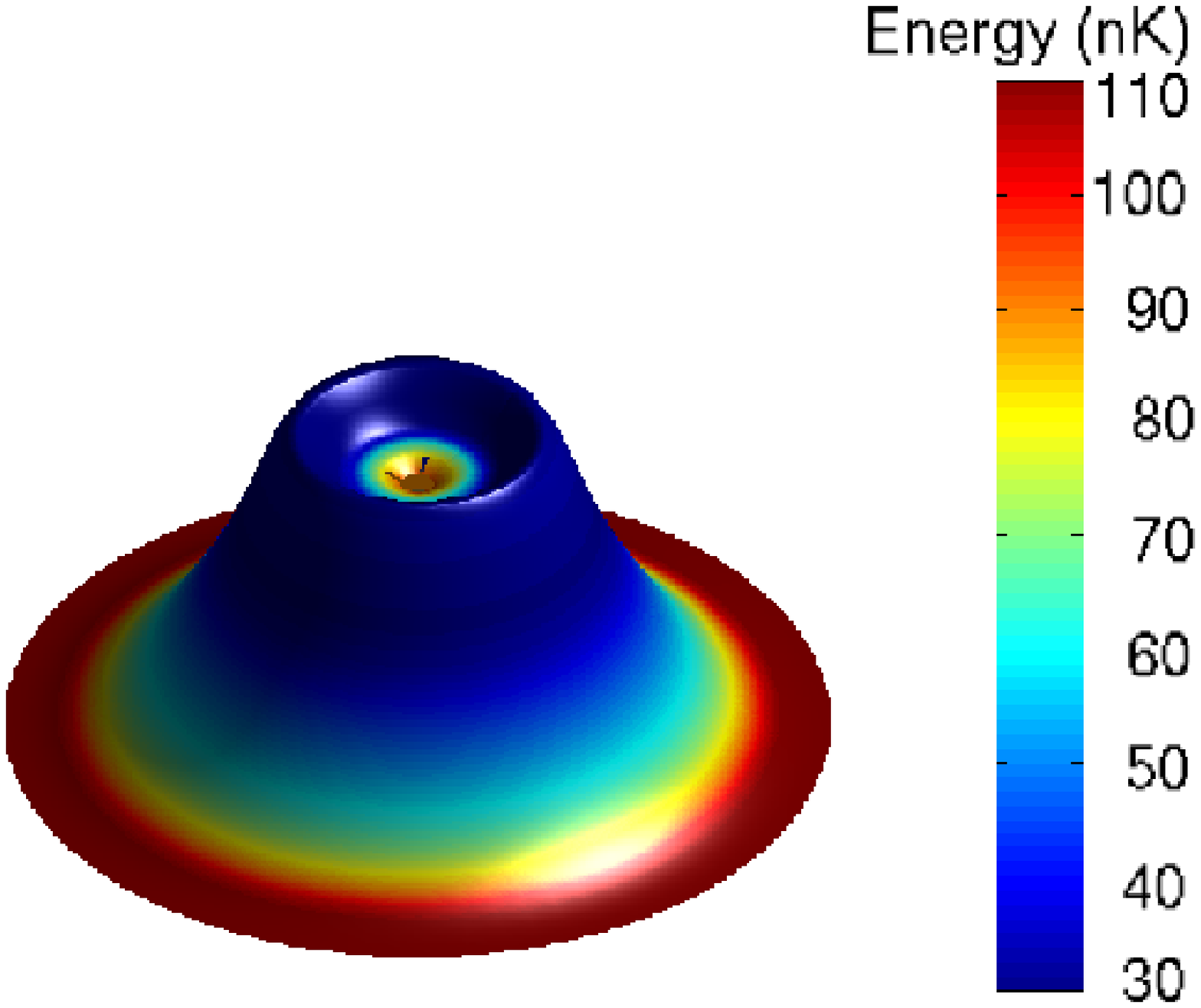}
\end{center}
\caption{\label{ripple}
  Density distribution of the noncondensate atoms near the end of the
  simulation (0.8~ms) on which we have superimposed the energy per
  particle as a colormap. The range of energies, of order 100~nK, is
  consistent with the characteristic scale of the burst particle
  energies in the Bosenova experiment. Note that hot atoms are
  generated in the center of the cloud during the atom-molecule
  oscillations since this is where the atom-molecule coupling is
  strongest (the coupling strength varies as the square root of the
  density). As the hot particles radiate outward a ring can be
  observed.}
\end{figure}

Obviously an important quantity to calculate for these expanding
noncondensed atoms is the effective temperature, or energy per
particle, since this quantity is observed experimentally. This is
illustrated in Figure~\ref{ripple} where superimposed on an
illustration of the density is a colormap of the temperature. The
hottest atoms generated in the center of the cloud are of comparable
energy scale to that seen in the experiment, being on the order of
100~nK.

\section{Conclusion}

We emphasize that the work presented here is a model calculation to
illustrate the feasibility for the burst to be generated through
atom-molecule coupling. However, there are a number of important
distinctions with the experimental situation which would have to be
accounted for before making a direct comparison. These simulations
contain no inelastic three-body loss and particle number is absolutely
conserved. In reality three-body loss may be important to the
experiment, but we suggest with this work that three-body loss is not
the only mechanism for producing a noncondensed burst during the
collapse.

It should be emphasized that if our hypothesis for the burst
generation is correct, the noncondensate atoms that are produced by
this mechanism are not simply generated in a thermal component, but
are instead generated in a fundamentally intriguing quantum state.
The process of dissociation of molecules into atom pairs produces
macroscopic correlations reminiscent of a squeezed vacuum state in
quantum optics. This means that every atom in the burst with momentum
$\boldvec k$ would have an associated partner with momentum $-\boldvec
k$. In principle, the correlations could be directly observed in
experiments through coincidence measurements providing clear evidence
as to whether this is the dominant mechanism for the burst generation
in the Bosenova.

\section{Acknowledgements}

We would like to thank Servaas Kokkelmans and Lincoln Carr for
discussions. M.H. and J.M. were supported by the U.S. Department of
Energy, Office of Basic Energy Sciences via the Chemical Sciences,
Geosciences and Biosciences Division, and C.M. by the National Science
Foundation and by the Divisione
Cooperazione e Mobilit\`a Internazionale of the University of Trento.


\section*{References}
\begin{thebibliography}{}
  
\bibitem{donley}Donley E A, Claussen N R, Cornish S L, Roberts J L,
  Cornell E A and Wieman C E 2001 {\it Nature} {\bf 412} 295
  
\bibitem{duine} Duine R A and Stoof H T C 2001 {\it Phys. Rev. Lett.}
  {\bf 86} 2204 
  
\bibitem{santos} Santos L and Shlyapnikov G V 2002 {\it Phys. Rev. A.}
  {\bf 66} 011602
  
\bibitem{saito} Saito H and Ueda M 2002 {\it Phys. Rev. A} {\bf 65}
  033624 
  
\bibitem{savage} Savage C M, Robins N P and Hope J J 2003 {\it Phys.
    Rev. A } {\bf 67} 014304
  
\bibitem{jeon}Jeon G S, Yin L, Rhee S W and Thouless D J 2002 {\it
    Phys. Rev. A} {\bf 66} 011603 
  
\bibitem{donley2}Donley E A, Claussen N R, Thompson S T and Wieman C E
  2002 {\it Nature} {\bf 417} 529

\bibitem{kokkelmans1} Kokkelmans S J J M F and Holland M J 2002 {\it
    Phys. Rev. Lett.} {\bf 89} 180401
  
\bibitem{mackie} Mackie M, Suominen K A and Javanainen J 2002 {\it
    Phys. Rev. Lett.} {\bf 89} 180403
  
\bibitem{kohler} K\"ohler T, Gasenzer T and Burnett K 2003 {\it Phys.
    Rev. A} {\bf 67} 013601
  
\bibitem{chiofalo} Chiofalo M L, Kokkelmans S J J M F, Milstein J N
  and Holland M J 2002 {\it Phys. Rev. Lett.} {\bf 88} 090402
  
\bibitem{kokkelmans2}Kokkelmans S J J M F, Milstein J N, Chiofalo M L,
  Walser R and Holland M J 2002 {\it Phys. Rev. A} {\bf 65} 053617
  
\end{thebibliography}
\end{document}